\documentclass[11pt,a4paper]{article}
\pdfoutput=1

\usepackage[T1]{fontenc}
\usepackage{graphicx, subfig, caption, a4wide, fancyvrb, upquote}
\graphicspath{{Figures/}{./}}
\usepackage{longtable, multirow, enumitem}
\usepackage{amsmath, amsfonts}
\usepackage{chicago}

\usepackage[round]{natbib}
\usepackage[margin=1in]{geometry}
\usepackage[scaled=0.9]{inconsolata}  
\DefineVerbatimEnvironment{Code}{Verbatim}{baselinestretch=0.9}
\DefineVerbatimEnvironment{CodeInput}{Verbatim}{baselinestretch=0.9,fontshape=sl}
\DefineVerbatimEnvironment{CodeOutput}{Verbatim}{baselinestretch=0.9}

\makeatletter
\newcommand\code{\bgroup\@makeother\_\@makeother\~\@makeother\$\@codex}
\def\@codex#1{{\normalfont\ttfamily\hyphenchar\font=-1 #1}\egroup}
\makeatother

\let\proglang=\textsf
\newcommand{\pkg}[1]{{\fontseries{b}\selectfont #1}}

\newcommand{\doi}[1]{\href{http://dx.doi.org/#1}{\normalfont\texttt{doi:#1}}}

\usepackage{afterpage}
\newcommand{\T}{^{\top}}
\newcommand{\X}{\boldsymbol{X}}
\newcommand{\x}{\boldsymbol{x}}
\newcommand{\D}{\boldsymbol{D}}
\newcommand{\A}{\boldsymbol{A}}
\newcommand{\mub}{\boldsymbol{\mu}}
\newcommand{\Sigmab}{\boldsymbol{\Sigma}}
\newcommand{\betab}{\boldsymbol{\beta}}
\newcommand{\I}{\boldsymbol{I}}
\newcommand{\Omegab}{\boldsymbol{\Omega}}
\newcommand{\0}{\boldsymbol{0}}
\DeclareMathOperator{\diag}{diag} 
\newcommand{\BIC}{\text{BIC}}
\newcommand{\BICdiff}{\BIC_{\text{diff}}}
\newcommand{\BICclust}{\BIC_{\text{clust}}}
\newcommand{\BICnotclust}{\BIC_{\text{not\,clust}}}
\newcommand{\BICreg}{\BIC_{\text{reg}}}
\newcommand{\Xset}{\mathcal{X}}
\newcommand{\Xclust}{\mathcal{X}_{\text{clust}}}
\newcommand{\Xother}{\mathcal{X}_{\text{other}}}

\usepackage{color}
\definecolor{dodgerblue2}{RGB}{28,134,238}
\definecolor{red3}{RGB}{205,0,0}
\definecolor{green3}{RGB}{0,205,0}
\definecolor{grey}{RGB}{127,127,127}
\definecolor{gray}{gray}{0.3}
\definecolor{darkblue}{rgb}{0,0,0.4}
\definecolor{darkgreen}{rgb}{0,0.6,0}
\definecolor{darkred}{rgb}{0.6,0.0,0}

\begin{document}
	
\title{\pkg{clustvarsel}: A Package Implementing Variable Selection for Model-based Clustering in R}
\author{%
Luca Scrucca \\ Universit\`a degli Studi di Perugia \and
Adrian E. Raftery \\ University of Washington
\thanks{Luca Scrucca is Assistant Professor of Statistics, Dipartimento di Economia, Universit\`a degli Studi di Perugia, Via A. Pascoli, 20, 06123 Perugia, Italy; Email: luca@stat.unipg.it.
Adrian E. Raftery is Professor of Statistics and Sociology, Department of Statistics, Box 354322, University of Washington, Seattle, WA 98195-4322, USA; Email: raftery@u.washington.edu.
Raftery's research was supported by the Eunice Kennedy Shriver National Institute of Child Health and Development through grants nos. R01 HD054511 and R01 HD070936, and by a Science Foundation Ireland E.T.S.~Walton visitor award, grant reference 11/W.1/I2079. 
Raftery thanks Nial Friel and the School of Mathematical Sciences at University College Dublin for hospitality during the preparation of this paper. 
The authors thank Nema Dean for helpful discussions.} }
\date{\today}
\maketitle

\begin{abstract}
Finite mixture modelling provides a framework for cluster analysis based on parsimonious Gaussian mixture models.
Variable or feature selection is of particular importance in situations where only a subset of the available variables provide clustering information.
This enables the selection of a more parsimonious model, yielding more efficient estimates, a clearer interpretation and, often, improved clustering partitions. This paper describes the R package \pkg{clustvarsel} which performs subset selection for model-based clustering.
An improved version of the methodology of Raftery and Dean (2006) is implemented in the new version 2 of the package to find the (locally) optimal subset of variables with group/cluster information in a dataset. Search over the solution space is performed using either a stepwise greedy search or a headlong algorithm. Adjustments for speeding up these algorithms are discussed, as well as a parallel implementation of the stepwise search.
Usage of the package is presented through the discussion of several data examples. \\

\end{abstract}

\newpage
\baselineskip=18pt

\section{Introduction}

Cluster analysis is the search for a priori unknown group structure in data. 
Model-based clustering is increasingly becoming one of the most popular cluster analysis methods. Model-based clustering is based on Finite mixture models \citep{McLachlan:Peel:2000}, with each component density usually representing a cluster. For continuous data, Gaussian components are usually used to model clusters.
Model-based clustering as implemented in the \proglang{R} package \pkg{mclust} \citep{Fraley:Raftery:Murphy:Scrucca:2012} allows for automatic selection of the number of components, and selection of parsimonious covariance structures.

In cluster analysis, as in classification or other supervised learning tasks, the inclusion of noise variables, i.e. features without useful group information, can severely degrade the final results. 
In fact, the presence of noise variables can negatively impact both the estimation of the number of clusters in the data and the recovery of those groups. 
The new \pkg{clustvarsel} (version $\ge 2.0$) \proglang{R} package implements a wrapper method for automatic variable selection in model-based clustering (as implemented in the \pkg{mclust} package). 
Thus, the addition of the \pkg{clustvarsel} package allows for automatic variable selection to be included in the estimation process. 

\citet{Raftery:Dean:2006} introduced a stepwise variable selection methodology tailored to model-based clustering. An earlier version of \pkg{clustvarsel} (version 1) implemented this methodology. Variables designated as noise variables in this process were not required to be independent of the clustering variables. However, noise variables could be conditionally independent of the clustering, but still linearly dependent on the clustering variables. This linear dependency was modelled using linear regression. 

\citet{Maugis:etal:2009a} extended the framework of \citet{Raftery:Dean:2006} by allowing the noise variables to depend on a (possibly null) subset of the clustering variables via stepwise variable selection in the linear regression \citep[see also][]{Maugis:etal:2009b}. This allows for a more parsimonious modelling of the relationship between the noise variables and the clustering variables. For more detail on the variable selection framework from both \citet{Raftery:Dean:2006} and \citet{Maugis:etal:2009a} see Section \ref{sec:methods}. 

Section \ref{sec:clustvarsel} introduces the main function in the \pkg{clustvarsel} package, and discusses the options for the available arguments. In Section \ref{sec:examples}, several examples are presented by applying the methodology to both synthetic and real world datasets. 
Algorithmic speedups are discussed in Section \ref{sec:speed}, including a description of a parallel implementation of the stepwise greedy search in Section \ref{sec:parallel}. The paper concludes with some discussion and final remarks in Section \ref{sec:discuss}.

\section{Methodology}
\label{sec:methods}

Model-based clustering assumes that the observed data are generated from a mixture of $G$ components, each representing the probability distribution for a different group or cluster \citep{McLachlan:Peel:2000, Fraley:Raftery:2002}.
For continuous data, the density of each mixture component can be described by the multivariate Gaussian distribution. Thus, the general form of a Gaussian finite mixture model is 
$$
f(\x) = \sum_{g=1}^G \pi_g \phi(\x|\mub_g,\Sigmab_g),
$$ 
where $\pi_g$ represents the mixing probabilities, so that $\pi_g > 0$ and $\sum_{g=1}^G\pi_g=1$, $\phi(\cdot)$ is the multivariate Gaussian density with parameters $(\mub_g,\Sigmab_g)$ ($g=1,\ldots,G$).
Clusters are ellipsoidal, centred at the mean vector $\mub_g$, with other geometric features, such as volume, shape and orientation, determined by $\Sigmab_g$.
Parsimonious parameterisation of covariance matrices is available through the eigenvalue decomposition $\Sigmab_g = \lambda_g \D_g \A_g \D\T_g$, where $\lambda_g$ is a scalar controlling the volume of the ellipsoid, $\A_g$ is a diagonal matrix specifying the shape of the density contours, and $\D_g$ is an orthogonal matrix which determines the orientation of the corresponding ellipsoid \citep{Banfield:Raftery:1993, Celeux:Govaert:1995}. \citet[Table~1]{Fraley:Raftery:Murphy:Scrucca:2012} report some parameterisation of within-group covariance matrices available in the \proglang{R} package \pkg{mclust}, and the corresponding geometric characteristics.
% Maximum likelihood estimates for this type of mixture models can be computed via the EM algorithm \citep{Dempster:Laird:Rubin:1977, McLachlan:Krishnan:1996}. A recent survey of finite mixture modelling for clustering is contained in \citet{Melnykov:Maitra:2010}.

\citet{Raftery:Dean:2006} discussed the problem of variable selection for model-based clustering by recasting the problem as a model selection procedure.
Their proposal is based on the use of BIC to approximate Bayes factors to compare mixture models fitted on nested subsets of variables. A generalisation of their approach was later discussed by \citet{Maugis:etal:2009a, Maugis:etal:2009b}.

Let us suppose that the set of available variables, $\Xset$, is partitioned into three disjoint parts: the set of previously selected variables, $\Xclust$, the variable under consideration for inclusion or exclusion from the active set, $X_i$, and the set of the remaining variables, $\Xother \equiv \Xset \setminus \{ \Xclust \cup X_i\}$.

\cite{Raftery:Dean:2006} showed that the inclusion (or exclusion) of variables can be assessed using the following BIC difference:
\begin{equation}
\BICdiff = \BICclust(\Xclust,X_i) - \BICnotclust(\Xclust,X_i),
\label{eq:BICdiff}
\end{equation}
where $\BICclust(\Xclust,X_i)$ is the BIC value for the ``best'' clustering mixture model (i.e. assuming $G \ge 2$) fitted using the features set $\{\Xclust \cup X_i\}$, whereas $\BICnotclust(\Xclust,X_i)$ is the BIC value for no clustering for the same set of variables. 
The latter can be written as
\begin{equation}
\BICnotclust(\Xclust,X_i) = \BICclust(\Xclust) + \BICreg(X_i|\Xclust),
\label{eq:BICnotclust}
\end{equation}
i.e. the BIC value for the ``best'' clustering model fitted using the set $\Xclust$ plus the BIC value for the regression of the candidate  variable $X_i$ on the variables included in the set $\Xclust$.

The difference in BIC score \eqref{eq:BICdiff} is an approximation of the log of the Bayes factor comparing the model where the variable under consideration, $X_i$, is a clustering variable with the model where the variable is conditionally independent of the clustering. Large, positive values of $\BICdiff$ can be taken as an evidence that variable $X_i$ is useful for clustering. 

In all clustering models, the ``best'' model is identified with respect to the number of mixture components (assuming $G \ge 2$) and to model parameterisations. 
In the linear regression model term, $X_i$ can depend on all the variables in $\Xclust$, a subset of them, or none (complete independence).
Finally, note that in both equations \eqref{eq:BICdiff} and \eqref{eq:BICnotclust} the set of remaining variables, $\Xother$, plays no role.
For more details of the methodology see \citet{Raftery:Dean:2006}, and for improvements due to the subset selection in the regression model see \citet{Maugis:etal:2009a}.

Practical implementation of the above methodology requires the use of an algorithm for checking single variables for inclusion/exclusion from the set of selected clustering variables. 
A \textit{stepwise greedy search} algorithm checks 
the inclusion of each single variable not currently selected into the current 
set of selected clustering variables at each step. 
The variable that has the highest evidence of inclusion is proposed and, if its clustering evidence is stronger than the evidence against clustering, it is included. 
At every exclusion step, the algorithm checks the removal of each single variable in the currently selected set of clustering variables, and proposes the variable that has the lowest evidence of clustering. The proposed variable is removed if the evidence for its being a clustering variable is weaker than the 
evidence against.
This is similar to the idea for stepwise regression and may suffer from the same instabilities mentioned in \citet{Miller:1990} inherent in that approach (although this has not been apparent in the simulations and examples tried thus far).

The stepwise algorithm can be implemented in a \textit{forward/backward} fashion, i.e. starting from the empty set of clustering variable and then continuing to add or remove features until there is no evidence of further clustering variables. It can also be implemented in a \textit{backward/forward} fashion, i.e. starting from the full set of features as clustering variables and then continuing to remove or add features until there is no evidence of further clustering variables.

Another possible algorithm is the \textit{headlong search}, which involves potentially checking less variables at each inclusion or exclusion step, and so may be quicker than the stepwise greedy search (at a possible price in terms of performance) for use on datasets with a large number of variables. At each inclusion step, the headlong algorithm only checks single variables not currently in the set of clustering variables until the difference between the BIC for clustering versus not clustering is above a pre-specified upper level (a default value of 0 implies that the evidence for clustering is greater than that for not clustering). 

Because the algorithm stops once this criterion is satisfied, it will not necessarily check all the variables available, and the variable selected will not necessarily be the best possible feature at that step. Any variables that are checked during this step, whose difference between the BIC for clustering versus not clustering is below a pre-specified lower level, are removed from consideration for the rest of the algorithm (a default value of $-10$ means that there is a strong evidence against clustering). Because of this, possibly irrelevant variables can be removed early on in the algorithm and further reduce the number of variables checked at each step. 

Similarly, at each exclusion, the algorithm only checks single variables currently in the set of clustering variables until the BIC difference for clustering versus not clustering is below the pre-specified upper level. The algorithm stops checking once a variable satisfies this criterion, and that variable is removed from the set. If the difference in BIC is smaller than the lower level, then the variable is removed from consideration for the rest of the algorithm, otherwise it can still be checked in future inclusion/exclusion steps. 
See \citet{Badsberg:1992} for full details about the headlong algorithm.

\section[The R package clustvarsel]{The \proglang{R} package \pkg{clustvarsel}}
\label{sec:clustvarsel}

The \pkg{clustvarsel} package can be used to find the (locally) optimal subset of variables with group/cluster information in a dataset with continuous variables. In this section, usage of the main function \code{clustvarsel} and its arguments is described. 

The \pkg{clustvarsel} package depends on other packages available on CRAN for model fitting (\pkg{mclust}, \citet{Rpkg:mclust}), or for providing some facilities, such as parallelisation (\pkg{parallel}, \citet{Rpkg:parallel}; \pkg{doParallel}, \citet{Rpkg:doParallel}; \pkg{foreach}, \citet{Rpkg:foreach}; \pkg{iterators}, \citet{Rpkg:iterators}), and subset selection in regression models (\pkg{BMA}, \citet{Rpkg:BMA}). By loading the package as usual with \code{library(clustvarsel)}, it will also take care of making all the other packages available for the current session.

Once the \pkg{clustvarsel} package has been loaded, the main function a user needs to invoke is the following:
\begin{quote}
\begin{Code}
clustvarsel(data, 
            G = 1:9, 
            search = c("greedy", "headlong"),
            direction = c("forward", "backward"),
            emModels1 = c("E", "V"), 
            emModels2 = mclust.options("emModelNames"),
            samp = FALSE, 
            sampsize = round(nrow(data)/2), 
            hcModel = "VVV", 
            allow.EEE = TRUE, 
            forcetwo = TRUE,
            BIC.diff = 0, 
            BIC.upper = 0, 
            BIC.lower = -10, 
            itermax = 100, 
            parallel = FALSE)
\end{Code}
\end{quote}

The available arguments are:
\begin{description}[itemsep=1ex, parsep=0pt]

\item[\code{data}] A numeric matrix or data frame where rows correspond to observations and columns correspond to variables. Categorical variables are not allowed.

\item[\code{G}] An integer vector specifying the numbers of mixture components (clusters) for which the BIC is to be calculated. The default is \code{G = 1:9}.

\item[\code{search}] A character vector indicating whether a \code{"greedy"} or, potentially quicker but less optimal, \code{"headlong"} algorithm is to be used in the search for clustering variables.

\item[\code{direction}] A character vector indicating the type of search: \code{"forward"} starts from the empty model and at each step of the algorithm adds/removes a variable until the stopping criterion is satisfied; \code{"backward"} starts from the model with all the available variables and at each step of the algorithm removes/adds a variable until the stopping criterion is satisfied. For the \code{"headlong"} search only the \code{"forward"} algorithm is available.

\item[\code{emModels1}] A vector of character strings indicating the models to be fitted in the EM phase of univariate clustering. Possible models are \code{"E"} and \code{"V"}, described in \code{help(mclustModelNames)}.

\item[\code{emModels2}] A vector of character strings indicating the models to be fitted in the EM phase of multivariate clustering. Possible models are those described in \code{help(mclustModelNames)}.

\item[\code{samp}] A logical value indicating whether or not a subset of observations should be used in the hierarchical clustering phase used to get starting values for the EM algorithm.

\item[\code{sampsize}] The number of observations to be used in the hierarchical clustering subset. By default, a random sample of approximately half of the sample size is used.

\item[\code{hcModel}] A character string specifying the model to be used in hierarchical clustering for choosing the starting values used by the EM algorithm. By default, the unconstrained \code{"VVV"} covariance structure is used.

\item[\code{allow.EEE}] A logical value indicating whether a new clustering will be run with equal within-cluster covariance for hierarchical clustering to get starting values, if the clusterings with variable within-cluster covariance for hierarchical clustering do not produce any viable BIC values.

\item[\code{forcetwo}] A logical value indicating whether at least two variables will be forced to be selected initially, regardless of whether BIC evidence suggests bivariate clustering or not.

\item[\code{BIC.diff}] A numerical value indicating the minimum BIC difference between clustering and no clustering used to accept the inclusion of a variable in the set of clustering variables in a forward step of the greedy search algorithm. Furthermore, minus \code{BIC.diff} is used to accept the exclusion of a selected variable from the set of clustering variable in a backward step of the greedy search algorithm. Default is $0$.

\item[\code{BIC.upper}] A numerical value indicating the minimum BIC difference between clustering and no clustering used to select a clustering variable in the headlong search. Default is $0$.

\item[\code{BIC.lower}] A numerical value indicating the level of BIC difference between clustering and no clustering below which a variable will be removed from consideration in the headlong algorithm. Default is $-10$.

\item[\code{itermax}] An integer value giving the maximum number of iterations (of addition and removal steps) the selected algorithm is allowed to run for.

\item[\code{parallel}] This argument allows to specify if the selected \code{"greedy"} algorithm should be run sequentially or in parallel. The possible values are: 
\begin{enumerate}[itemsep=0pt, parsep=0pt, topsep=1ex, label=(\roman*)]
\item a logical value specifying if parallel computing should be used (\code{TRUE}) or not (\code{FALSE}, default) for running the algorithm;
\item a numerical value which gives the number of cores to employ (by default, this is obtained from function \code{detectCores} in the \pkg{parallel} package);
\item a character string specifying the type of parallelisation to use. The latter depends on system OS: on Windows OS only \code{"snow"} type functionality is available, whereas on Unix/Linux/Mac OSX both \code{"snow"} and \code{"multicore"} (default) functionalities are available.
\end{enumerate}
Options (ii) and (iii) imply that the search is performed in parallel. By default the algorithm is run sequentially.

\end{description}

A basic \code{clustvarsel} function call needs to input a matrix or data frame containing the data to analyse. Fine tuning is possible by specifying the arguments described above. The following section presents some examples of its usage in practice.

\section{Examples}
\label{sec:examples}

In this section we present some data analysis examples based on simulated data and on well-known real datasets.

% Luca: all the examples use mclust.options(hcUse = "SVD")
% which is automatically set by loading the clustvarsel
% package. This requires the latest version of both 
% \pkg{mclust} (>= 4.4) and \pkg{clustvarsel} (>= 2.1) 

\subsection{Simulated data}
\label{sec:simul}

%\cite{Maugis:etal:2009a, Maugis:etal:2009b} proposed a method based on Raftery and Dean's approach by allowing the irrelevant variables to be explained by only a relevant variable subset. They proposed a two-step nested backward stepwise algorithm, with the second inner backward algorithm which is needed to identify the block of variables independent from the clustering ones. 

We consider some of the synthetic data examples described in \citet{Maugis:etal:2009a}. Samples were simulated for a 10-dimensional feature vector where only the first two variables provide clustering information. These were generated from a mixture of four Gaussian distributions $\X_{[1:2]} \sim N(\mub_k, \I_2)$ with $\mub_1=(-2,-2)$, $\mub_2=(-2,2)$, $\mub_3 = -\mub_2$, $\mub_4 = -\mub_1$, and mixing probabilities $\pi = (0.3,0.2,0.3,0.2)$. The remaining eight variables were simulated according to the model $\X_{[3:10]} = \X_{[1:2]}\betab + \epsilon$, where $\epsilon \sim N(\0, \Omegab)$. Different settings for $\betab$ and $\Omegab$ define seven different scenarios \citep[see Table~1 in][]{Maugis:etal:2009a}. These range from independence of clustering variables on the other features (model 1 and 2) to cases of increasing degree of dependence of the irrelevant variables on the clustering ones (model 3 to 7). 
In the following, we focus only on some of the scenarios. For ease of reading, the values of the parameters for such scenarios are reported in Table~\ref{tab:maugisetal:scenarios}.

\begin{table}[htb]
\caption[Parameter settings for the scenarios used in the simulation study]
{Parameter settings for the scenarios used to generated synthetic data: $\betab$ defines the correlation of irrelevant variables on clustering variables, whereas $\Omegab$ is the covariance structure of the noise component. $\0_p$ indicates the $(2 \times p)$ matrix of zeroes, and $\I_p$ the $(p \times p)$ identity matrix.}
\label{tab:maugisetal:scenarios}
\centering
\begin{tabular*}{0.95\textwidth}{@{\extracolsep{\fill}}llcll}
\hline\noalign{\smallskip}
Scenario & Parameters && Scenario & Parameters \\
\hline\noalign{\smallskip}
\textit{Model 1} & $\betab = \0_8$ &&
\textit{Model 5} & 
$\betab = \left(\begin{matrix} 0.5 & 0 & 2 & 0\\ 0 & 1 & 0 & 3\end{matrix}\;\0_4\right)$ \\[2ex]
& $\Omegab = \I_8$ &&& 
$\Omegab = \diag(\I_2, 0.5\I_2, \I_4)$ \\[2ex]
\textit{Model 4} & 
$\betab =  \left(\begin{matrix} 0.5 & 0 \\ 0 & 1 \end{matrix}\;\0_6\right)$ &&
\textit{Model 7} & 
$\betab = \left(\begin{matrix} 0.5 & 0 & 2 & 0 & 2 & 0.5 & 2 & 0\\ 0 & 1 & 0 & 3  & 0.5 & 1 & 0 & 3\end{matrix}\right)$ \\[2ex]
& $\Omegab = \I_8$ &&& 
$\Omegab = \diag(\I_2, 0.5\I_4, \I_2)$ \\
\noalign{\smallskip}\hline
\end{tabular*}
\end{table}

The simulation results were evaluated using the following criteria:
\begin{itemize}[itemsep=0pt, parsep=0pt, topsep=1ex]
\item Variable Selection Error Rate (VSER) to assess \textit{variable selection performance}. VSER is defined as the ratio of the number of errors in selecting (or not selecting) variables to the total number of variables in the set. A perfect recovey of clustering variables gives VSER $= 0$, while VSER can be no greater than 1.
\item Adjusted Rand Index \citep[ARI,][]{Hubert:Arabie:1985} to measure \textit{classification accuracy}. A perfect classification gives ARI $= 1$, whereas ARI $= 0$ for a random classification.
% The number of selected variables is also reported. 
\end{itemize}

Table~\ref{tab:maugis:sim1} shows the results from a simulation study for the above synthetic data using sample sizes $n=200$ and $n=1000$. 
The methods compared are \code{MCLUST}, the best GMM using the full set of variables, \code{CLUSTVARSEL[fwd]} and \code{CLUSTVARSEL[bkw]}, the best GMM using the subset of relevant clustering variables selected by, respectively, the forward/backward and backward/forward greedy search, and \code{SPARSEKMEANS}, a sparse version of $k$-means algorithm \citep{Witten:Tibshirani:2010}. 
Because the last method needs the number of clusters to be fixed in advance, we also included in the comparison versions of the methods based on GMMs with the number of components fixed at the true number of clusters, i.e. $G=4$. 
Finally, note that true subset size is 2, so the optimal VSER should be 0, and the best average ARI value attainable, using the true clustering variables and fixed $G = 4$ components, is about $0.88$.

\begin{table}[htb]
\caption[Results of the simulation study based on the Maugis et al (2009) setup]
{Average values based on 100 simulation runs for some of the models in \citet{Maugis:etal:2009a}, where only two (out of ten) variables are relevant for clustering. The first four models use $G=4$ fixed number of clusters. \code{CLUSTVARSEL[fwd]} uses the forward/backward greedy search, whereas \code{CLUSTVARSEL[bkw]} employs the backward/forward greedy search. True subset size is 2. Smaller values of VSER and larger values of ARI are better. The best values for each experiment and each criterion are shown in bold font. For subset size, these are the values closest to 2, for VSER the smallest values, and for ARI the largest values.}
\label{tab:maugis:sim1}
\centering
\begin{tabular}{lrrrcrrr}
\hline\noalign{\smallskip}
Model & Subset size & VSER & ARI && Subset size & VSER & ARI \\
\hline\noalign{\smallskip}
%%%%%%%%%%%%%%%%%%%%%%%%%%%%%%%%%%%%%%%%%%%%%%%%%%%%%%%%%%
Scenario 1 & \multicolumn{3}{c}{$n=200$} && \multicolumn{3}{c}{$n=1000$} \\
\cline{2-4}\cline{6-8}
\code{MCLUST,G=4}           & 10.00 & .800 & .867 && 10.00 & .800 & .882 \\ 
\code{SPARSEKMEANS,G=4}     &  9.91 & .791 & {\bf .872} &&  9.37 & .737 & .881 \\ 
\code{CLUSTVARSEL[fwd],G=4} & {\bf 2.05} & {\bf .005} & {\bf .872} &&  {\bf 2.01} & {\bf .001} &{\bf .885} \\
\code{CLUSTVARSEL[bkw],G=4} & {\bf 2.05} & {\bf .005} & .873 &&  {\bf 2.01} & {\bf .001} &{\bf .885} \\  
\code{MCLUST}               & 10.00 & .800 & .770 && 10.00 & .800 & .882 \\ 
\code{CLUSTVARSEL[fwd]}     &  {\bf 2.05} & {\bf .005} & {\bf .872} &&  {\bf 2.01} & {\bf .001} &{\bf .885} \\  
\code{CLUSTVARSEL[bkw]}     &  3.86 & .278 & .681 &&  {\bf 2.01} & {\bf .001} &{\bf .885} \\ 
\hline\noalign{\smallskip}
%%%%%%%%%%%%%%%%%%%%%%%%%%%%%%%%%%%%%%%%%%%%%%%%%%%%%%%%%%
Scenario 4 & \multicolumn{3}{c}{$n=200$} && \multicolumn{3}{c}{$n=1000$} \\
\cline{2-4}\cline{6-8}
\code{MCLUST,G=4}           & 10.00 & .800 & .828 && 10.00 & .800 & .881 \\
\code{SPARSEKMEANS,G=4}     &  9.89 & .789 & .834 &&  9.35 & .735 & .842 \\ 
\code{CLUSTVARSEL[fwd],G=4} &  {\bf 2.03} & {\bf .003} & {\bf .881} &&  {\bf 2.01} & {\bf .001} & {\bf .886} \\ 
\code{CLUSTVARSEL[bkw],G=4} &  {\bf 2.03} & {\bf .003} & {\bf .881} &&  {\bf 2.01} & {\bf .001} & {\bf .886} \\ 
\code{MCLUST}               & 10.00 & .800 & .698 && 10.00 & .800 & .881 \\
\code{CLUSTVARSEL[fwd]}     &  2.05 & .007 & .877 &&  {\bf 2.01} & {\bf .001} & {\bf .886} \\ 
\code{CLUSTVARSEL[bkw]}     &  2.98 & .162 & .752 &&  {\bf 2.01} & {\bf .001} & {\bf .886} \\  
\hline\noalign{\smallskip}
%%%%%%%%%%%%%%%%%%%%%%%%%%%%%%%%%%%%%%%%%%%%%%%%%%%%%%%%%%
Scenario 5 & \multicolumn{3}{c}{$n=200$} && \multicolumn{3}{c}{$n=1000$} \\
\cline{2-4}\cline{6-8}
\code{MCLUST,G=4}           & 10.00 & .800 & .847 && 10.00 & .809 & .879 \\ 
\code{SPARSEKMEANS,G=4}     &  9.34 & .736 & .801 &&  7.09 & .509 & .857 \\ 
\code{CLUSTVARSEL[fwd],G=4} &  {\bf 2.00} & {\bf .014} & {\bf .881} &&  {\bf 2.01} & {\bf .001} & {\bf .884} \\ 
\code{CLUSTVARSEL[bkw],G=4} &  2.03 & .035 & .880 &&  2.03 & .003 & .884 \\ 
\code{MCLUST}               & 10.00 & .800 & .461 && 10.00 & .800 & .879 \\ 
\code{CLUSTVARSEL[fwd]}     &  1.99 & .017 & .873 &&  {\bf 2.01} & {\bf .001} & {\bf .884} \\ 
\code{CLUSTVARSEL[bkw]}     &  2.06 & .030 & .868 &&  2.03 & .003 & {\bf .884} \\ 
\hline\noalign{\smallskip}
%%%%%%%%%%%%%%%%%%%%%%%%%%%%%%%%%%%%%%%%%%%%%%%%%%%%%%%%%%
Scenario 7 & \multicolumn{3}{c}{$n=200$} && \multicolumn{3}{c}{$n=1000$} \\
\cline{2-4}\cline{6-8}
\code{MCLUST,G=4}           & 10.00 & .800 & .838 && 10.00 & .800 & .880 \\  
\code{SPARSEKMEANS,G=4}     &  9.04 & .704 & .847 &&  9.00 & .700 & .865 \\ 
\code{CLUSTVARSEL[fwd],G=4} &  {\bf 2.00} & .032 & {\bf .874} &&  {\bf 2.00} & {\bf .000} & {\bf .885} \\ 
\code{CLUSTVARSEL[bkw],G=4} &  2.03 & .021 & .872 &&  2.01 & .003 & {\bf .885} \\ 
\code{MCLUST}               & 10.00 & .800 & .447 && 10.00 & .800 & .880 \\  
\code{CLUSTVARSEL[fwd]}     &  {\bf 2.00} & .024 & {\bf .874} &&  {\bf 2.00} & {\bf .000} & {\bf .885} \\ 
\code{CLUSTVARSEL[bkw]}     &  2.02 & {\bf .014} & .868 &&  2.01 & .003 & {\bf .885} \\ 
%%%%%%%%%%%%%%%%%%%%%%%%%%%%%%%%%%%%%%%%%%%%%%%%%%%%%%%%%%
\hline\noalign{\smallskip}
\multicolumn{8}{l}{\footnotesize Standard errors for VSER and ARI are all $\le .030$.}
\end{tabular}
\end{table}

Compared to the performance of \code{CLUSTVARSEL} reported in Table~1 of \citet{Maugis:etal:2009a}, the new version of the algorithm is able to correctly discard irrelevant variables, both when they are independent of the clustering ones and when they are correlated. 

When $G$ is fixed at the true number of clusters, \code{MCLUST} gives slightly less accurate results for $n=200$, except in the case of complete independence (scenario 1). \code{CLUSTVARSEL} provides equivalent accuracy, both if a forward/backward search or a backward/forward search is used. 
\code{SPARSEKMEANS} shows results equivalent to greedy search in term of accuracy, but it tends to select (i.e. assigns weights different from zero to) too many variables. Consequently, the VSER of \code{SPARSEKMEANS} is always worse than that of \code{CLUSTVARSEL}.

When $G$ is unknown, \code{MCLUST} often provides inaccurate clustering, in particular when $n=200$. On the contrary, \code{CLUSTVARSEL} is generally able to select the true clustering variables (i.e. VSER is near or exactly zero), and also provides very accurate clustering (i.e. ARI is close to $0.88$). The only exceptions are scenarios 1 and 4, for the  backward/forward search when $n=200$. In these cases the number of selected variables is slightly larger, which in turn causes a small degradation of clustering accuracy. However, for $n=1,000$ the forward/backward and backward/forward greedy searches are equivalent.

\afterpage{\clearpage}
\subsection{Crabs data}

The crabs dataset in the \pkg{MASS} package contains five morphological measurements on 200 specimens of Leptograpsus variegatus crabs recorded on the shore in Western Australia \citep{Campbell:Mahon:1974}. Crabs are classified according to their color (blue and orange) and sex, giving four groups. Fifty specimens are available for each combination of colour and sex.
\begin{CodeInput}
R> data(crabs, package = "MASS")
R> X = crabs[,4:8]
R> Class = with(crabs, paste(sp, sex, sep = "|"))
R> table(Class)
\end{CodeInput}
\begin{CodeOutput}
Class
B|F B|M O|F O|M 
 50  50  50  50 
\end{CodeOutput}

First we look at the result obtained using the function \code{Mclust} from the \pkg{mclust} package, with the best model selected by BIC for clustering on all the variables, allowing all possible parameterisations and the number of groups to range over 1 to 5:
% mclust.options(hcUse = "PCS")
\begin{CodeInput}
R> mod1 = Mclust(X, G = 1:5)
R> summary(mod1)
\end{CodeInput}
\begin{CodeOutput}
----------------------------------------------------
Gaussian finite mixture model fitted by EM algorithm 
----------------------------------------------------

Mclust EEV (ellipsoidal, equal volume and shape) model with 4 components:

 log.likelihood   n df       BIC      ICL
      -1241.006 200 68 -2842.298 -2854.29

Clustering table:
 1  2  3  4 
60 55 39 46 
\end{CodeOutput}

The estimated MAP classification is obtained from \code{mod1$classification}, so a table comparing the estimated and the true classifications, the corresponding misclassification error rate and the adjusted Rand index (ARI), can be obtained as follows:
\begin{CodeInput}
R> table(Class, mod1$classification)
\end{CodeInput}
\begin{CodeOutput}
Class  1  2  3  4
  B|F 49  0  0  1
  B|M 11  0 39  0
  O|F  0  5  0 45
  O|M  0 50  0  0
\end{CodeOutput}
\begin{CodeInput}
R> classError(Class, mod1$classification)$errorRate
\end{CodeInput}
\begin{CodeOutput}
[1] 0.085
\end{CodeOutput}
\begin{CodeInput}
R> adjustedRandIndex(Class, mod1$classification)
\end{CodeInput}
\begin{CodeOutput}
[1] 0.793786
\end{CodeOutput}

The algorithm for selecting the variables that are useful for clustering can be run with the following code:
\begin{CodeInput}
R> result = clustvarsel(X, G = 1:5)
R> result
\end{CodeInput}
\begin{CodeOutput}
'clustvarsel' model object:

Stepwise (forward) greedy search:
   Variable proposed        BIC  BIC difference  Type of step  Decision
1                 CW  -1408.710        -6.21775           Add  Accepted
2                 RW  -1908.964       127.38583           Add  Accepted
3                 FL  -2357.252        81.24626           Add  Accepted
4                 FL  -2357.252        81.24074        Remove  Rejected
5                 BD  -2609.777        56.08094           Add  Accepted
6                 BD  -2609.777        71.39446        Remove  Rejected
7                 CL  -2609.777       -31.07119           Add  Rejected
8                 BD  -2609.777        71.39446        Remove  Rejected

Selected subset: CW, RW, FL, BD
\end{CodeOutput}

By default, a greedy forward/backward search is used. 
The printed output shows the trace of the algorithm: 
at each step the most important variable is considered for addition or deletion from the set of clustering variables, with each proposal which can be accepted or rejected. In this case, the final subset contains four out of five morphological features:
\begin{CodeInput}
R> result$subset
\end{CodeInput}
\begin{CodeOutput}
CW RW FL BD 
 4  2  1  5 
\end{CodeOutput}

The same subset is also obtained by using a backward/forward greedy search:
\begin{CodeInput}
R> clustvarsel(X, G = 1:5, direction = "backward")
\end{CodeInput}
\begin{CodeOutput}
'clustvarsel' model object:

Stepwise (backward) greedy search:
   Variable proposed        BIC  BIC difference  Type of step  Decision
1                 CL  -2609.777       -31.07119        Remove  Accepted
2                 BD  -2609.777        56.08094        Remove  Rejected

Selected subset: FL, RW, CW, BD
\end{CodeOutput}

The identified subset can be used for fitting the final clustering model as follows:
\begin{CodeInput}
R> Xs = X[,result$subset]
R> mod2 = Mclust(Xs, G = 1:5)
R> summary(mod2)
\end{CodeInput}
\begin{CodeOutput}
----------------------------------------------------
Gaussian finite mixture model fitted by EM algorithm 
----------------------------------------------------

Mclust EEV (ellipsoidal, equal volume and shape) model with 4 components:

 log.likelihood   n df       BIC       ICL
      -1180.378 200 47 -2609.777 -2624.892

Clustering table:
 1  2  3  4 
53 60 40 47 
\end{CodeOutput}
The accuracy of the clustering obtained on the selected subset of variables is obtained as:
\begin{CodeInput}
R> table(Class, mod2$classification)
\end{CodeInput}
\begin{CodeOutput}
Class  1  2  3  4
  B|F  0 50  0  0
  B|M  0 10 40  0
  O|F  3  0  0 47
  O|M 50  0  0  0
\end{CodeOutput}
\begin{CodeInput}
R> classError(Class, mod2$classification)$errorRate
\end{CodeInput}
\begin{CodeOutput}
[1] 0.065
\end{CodeOutput}
\begin{CodeInput}
R> adjustedRandIndex(Class, mod2$classification)
\end{CodeInput}
\begin{CodeOutput}
[1] 0.8399679
\end{CodeOutput}

\subsection{Coffee data}

Data on twelve chemical constituents of coffee for 43 samples were collected from 29 countries around the world \citep{Streuli:1973}. Each coffee sample is either of the Arabica or Robusta variety. The dataset is available in the \proglang{R} package \pkg{pgmm}. 

\begin{CodeInput}
R> data(coffee, package = "pgmm")
R> X = as.matrix(coffee[,3:14])
R> Class = factor(coffee$Variety, levels = 1:2, labels = c("Arabica", "Robusta"))
R> table(Class)
\end{CodeInput}
\begin{CodeOutput}
Class
Arabica Robusta 
     36       7
\end{CodeOutput}

\begin{CodeInput}
R> mod1 = Mclust(X)
R> summary(mod1)
\end{CodeInput}
\begin{CodeOutput}
----------------------------------------------------
Gaussian finite mixture model fitted by EM algorithm 
----------------------------------------------------

Mclust VEI (diagonal, equal shape) model with 3 components:

 log.likelihood  n df       BIC       ICL
      -392.9397 43 52 -981.4619 -981.6379

Clustering table:
 1  2  3 
22 14  7 
\end{CodeOutput}

Model-based clustering applied to this dataset selects the \code{VEI} model with 3 components. 
The clustering table and the corresponding adjusted Rand index (ARI) are the following:
\begin{CodeInput}
R> table(Class, mod1$classification)
\end{CodeInput}
\begin{CodeOutput}
Class      1  2  3
  Arabica 22 14  0
  Robusta  0  0  7
\end{CodeOutput}
\begin{CodeInput}
R> adjustedRandIndex(Class, mod1$classification)
\end{CodeInput}
\begin{CodeOutput}
[1] 0.3833116
\end{CodeOutput}
The Arabica variety appears to be split into two sub-varieties, whereas the Robusta is correctly identified as a single cluster. As a result, a small value of ARI is obtained.

Now, we may try variable selection to drop irrelevant features, and see if we can improve upon the above model. 
The following code uses the backward/forward greedy search for variable selection, which by default is performed over all the covariance decomposition models and numbers of mixture components from 1 up to 9: 
\begin{CodeInput}
R> result = clustvarsel(X, direction = "backward")
R> result
\end{CodeInput}
\begin{CodeOutput}
'clustvarsel' model object:

Stepwise (backward) greedy search:
     Variable proposed        BIC  BIC difference  Type of step  Decision
1        Extract Yield  -788.3021      -10.930431        Remove  Accepted
2  Neochlorogenic Acid  -852.5413       -9.982637        Remove  Accepted
3     Chlorogenic Acid  -805.6227      -11.065351        Remove  Accepted
4        Extract Yield  -999.1101       -9.315106           Add  Rejected
5  Isochlorogenic Acid  -816.8139      -13.958685        Remove  Accepted
6        Extract Yield  -936.2985       66.325955           Add  Accepted
7        Extract Yield  -936.2985       66.325955        Remove  Rejected
8  Isochlorogenic Acid  -999.1101      -90.028182           Add  Rejected

Selected subset: 
Water, Bean Weight, ph Value, Free Acid, Mineral Content, Fat, Caffine, 
Trigonelline, Extract Yield
\end{CodeOutput}

Then, the clustering model estimated on the selected subset of variables is:
\begin{CodeInput}
R> mod2 = Mclust(X[,result$subset])
R> summary(mod2)
\end{CodeInput}
\begin{CodeOutput}
----------------------------------------------------
Gaussian finite mixture model fitted by EM algorithm 
----------------------------------------------------

Mclust EEI (diagonal, equal volume and shape) model with 3 components:

 log.likelihood  n df       BIC       ICL
      -443.2269 43 38 -1029.379 -1030.937

Clustering table:
 1  2  3 
22 14  7
\end{CodeOutput}
\begin{CodeInput}
R> table(Class, Cluster = mod2$class)
\end{CodeInput}
\begin{CodeOutput}
         Cluster
Class      1  2  3
  Arabica 22 14  0
  Robusta  0  0  7
\end{CodeOutput}
\begin{CodeInput}
R> table(Class, Cluster = mod2$class)
\end{CodeInput}
\begin{CodeOutput}
         Cluster
Class      1  2  3
  Arabica 22 14  0
  Robusta  0  0  7
\end{CodeOutput}

Both the covariance parameterisation (EEE) and the number of mixture components (3) used with the selected features subset agree with those from the model using all the variables. The final clustering confirms the structure we already discussed, in particular the two sub-varieties of Arabica coffee. 

\begin{figure}
\centering
\includegraphics[width=0.7\linewidth]{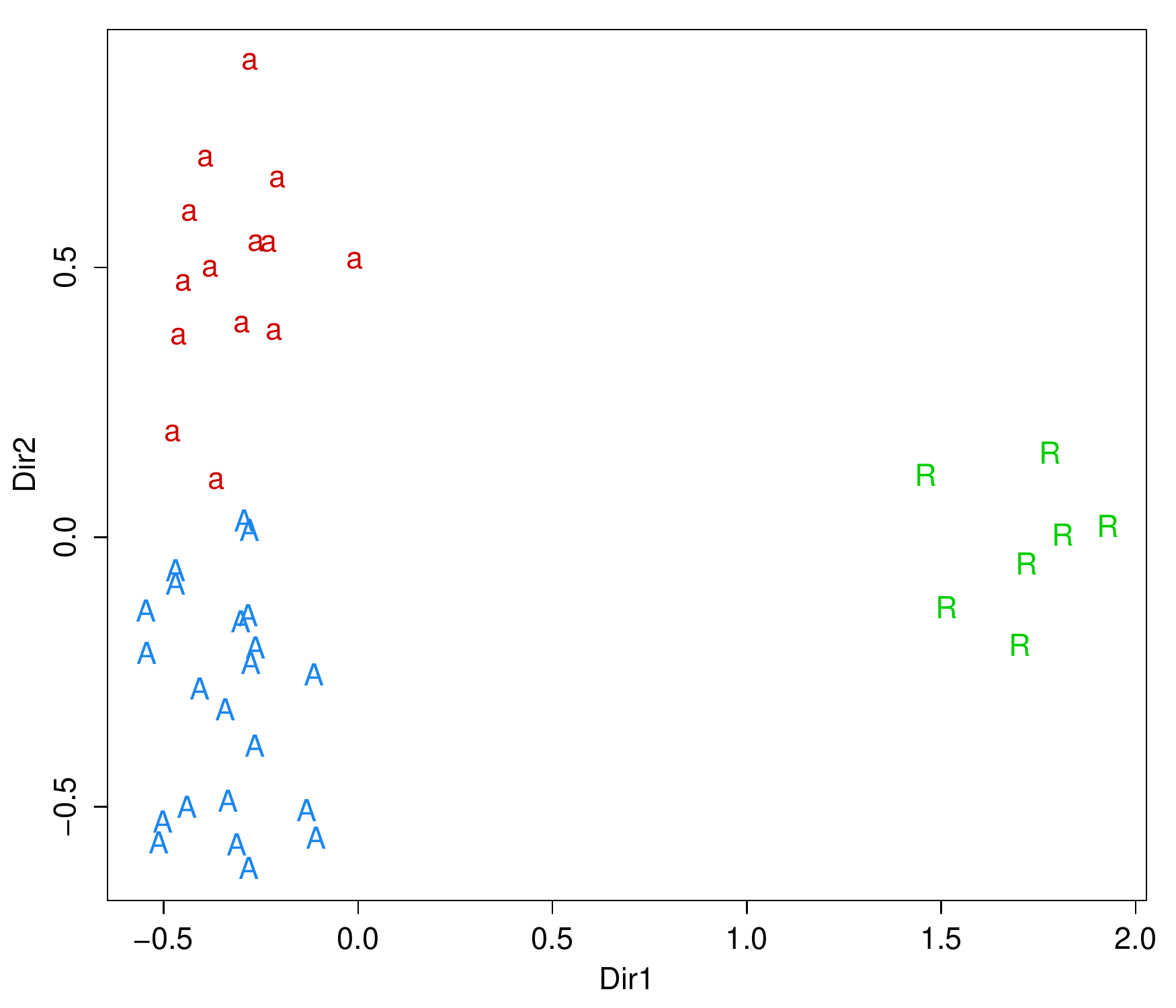}
\caption[Clustering of coffee data samples]
{Projection of coffee data samples marked according to the clustering obtained from the variables selected using the forward/backward greedy search. The symbol \textcolor{green3}{\sf R} indicates Robusta coffees, \textcolor{dodgerblue2}{\sf A} and \textcolor{red3}{\sf a} the sub-varieties of Arabica coffees.}
\label{fig1:coffee}
\end{figure}

To show graphically these findings, we may project the data onto a dimension reduced subspace by using the methodology described in \citet{Scrucca:2010}: 
\begin{CodeInput}
R> mod2dr = MclustDR(mod2)
R> plot(mod2dr, what = "scatterplot", symbols = c("A", "a", "R"))
\end{CodeInput}
% dev.copy2pdf(file = "~/Stat/clustvarsel/Paper/Figures/coffee_fig1.pdf", width = 7, height = 6)
From Figure~\ref{fig1:coffee} there is an evident separation between Arabica and Robusta coffee samples along the first direction. Moreover, it seems to confirm the non homogeneous group of Arabica samples, which splits in two sub-varieties along the second direction. 

\subsection{Simulated high-dimensional data}
\label{sec:highdim}

\citet[Sec. 3.3.2]{Witten:Tibshirani:2010} discussed an example where five clustering variables are conditionally independent given the cluster memberships, whereas the remaining twenty features are simply independent standard normal variables, also independent from the clustering ones.
The first five variables are distributed according to a spherical Gaussian distribution with mean $\mub_1=(\mu, \mu, \ldots, \mu)$, $\mub_2 = \0$, $\mub_3 = -\mub_1$, where $\mu=1.7$, and common unit standard deviation. We replicated this experiment (denoted by WT) with varying sample sizes ($n_g$ cases for each group) and for a set of different techniques. 

Table~\ref{tab:WTsim} reports the variable selection error rate (VSER) and the classification error rate (CER). 
As already mentioned in Section~\ref{sec:examples}.1, the VSER  is defined as the ratio of the number of errors in selecting (or not selecting) variables with respect to the total number of variables considered. 
The CER between two partitions, which is equivalent to one minus the Rand index \citep{Rand:1971}, is equal to 0 in the case of perfect agreement, and becomes larger for increasing disagreement \citep[for a formal definition see][]{Witten:Tibshirani:2010}.
Smaller values of both VSER and CER are better.
These two measures have been chosen for the purpose of comparison with the results in \citet[Tab. 4]{Witten:Tibshirani:2010} and \citet[Sect. 3.1]{Celeux:etal:2013}.

The model with uniformly better performance is the  \code{MCLUST} model with the first five variables, i.e. the model which most resembles the data generation mechanism. However, this model is not available when we do not know the 
clustering variables, as we are assuming here.
\code{SPARSEKMEANS} performs well in term of accuracy, but the number of selected variables increases with group sample size, and all the features are selected for $n_g=50$.
\code{MCLUST} using all the variables and EII parameterisation, both at the hierarchical initialisation step and for the mixture modelling, has quite good accuracy and improves as group sample size increases.
\code{CLUSTVARSEL} using backward/forward greedy search, with EII parameterisation both for modelling and initialisation also has good accuracy, and improves as group sample size gets larger. For $n_g=50$ it is equivalent to the best models. However, the VSER is better than for the other methods, and improves as $n_g$ gets larger. Thus, for increasing sample size it converges to the true subset size.
Note that forward/backward greedy search is clearly less accurate than the backward/forward search, with the VSER which is also larger, so the performance is overall worst than that of backward/forward search.
Finally, results for \code{CLUSTVARSEL} using backward/forward greedy search are essentially equivalent to those obtained with the \code{SELVARCLUSTINDEP} software of \citet{Maugis:etal:2009b}.

\begin{table}[htb]
\centering
\caption[Results of the simulation study based on the Witten and 
Tibshirani (2010) setup]
{Average values based on 100 simulation runs for the WT simulation scheme with group sample size $n_g$. All models assume the number of clusters $G=3$ to be known. For all the \code{MCLUST} models, including those fitted by the variables subset selection algorithm, the EII parameterisation was used for both the hierarchical initialization and the mixture modeling. \code{CLUSTVARSEL[fwd]} uses the forward/backward greedy search, whereas \code{CLUSTVARSEL[bkw]} uses the backward/forward greedy search. Smaller values of both VSER and CER are better.}
\label{tab:WTsim}
\begin{tabular}{lrrrcrrr}
\hline\noalign{\smallskip}
& \multicolumn{3}{c}{VSER} & & \multicolumn{3}{c}{CER} \\
Model & $n_g=10$ & $n_g=20$ & $n_g=50$ & & $n_g=10$ & $n_g=20$ & $n_g=50$ \\
\hline\noalign{\smallskip}
\code{K-MEANS}                            & .800 & .800 & .800 && .258 & .248 & .224 \\ 
\code{SPARSEKMEANS}                       & .438 & .678 & .800 && .070 & .066 & .055 \\ 
\code{MCLUST[}$X_1,\ldots,X_5$\code{]}    & .000 & .000 & .000 && .063 & .064 & .054 \\ 
\code{MCLUST[}$X_1,\ldots,X_{25}$\code{]} & .800 & .800 & .800 && .129 & .093 & .060 \\ 
\code{CLUSTVARSEL[fwd]}                   & .268 & .200 & .054 && .370 & .275 & .090 \\ 
\code{CLUSTVARSEL[bkw]}                   & .180 & .081 & .033 && .151 & .082 & .053 \\ 
\code{SELVARCLUSTINDEP}                   & .216 & .098 & .032 && .162 & .089 & .057 \\ 
\hline\noalign{\smallskip}
\multicolumn{8}{l}{\footnotesize Standard errors are all $< .030$.}
\end{tabular}
\end{table}

\section{Adjustments for speeding up the algorithm}
\label{sec:speed}

\subsection{Sub-sampling at hierarchical initialisation step}
\label{sec:subsample}

The EM algorithm is initialised in \pkg{mclust} using the partitions obtained from model-based agglomerative hierarchical clustering. Efficient numerical algorithms for approximately maximizing the classification likelihood with multivariate normal models have been discussed by \citet{Fraley:1998}. However, for datasets having a large number of observations this step can be computationally expensive.

When the number of observations is large, we may allow \code{clustvarsel} to use only a subset of the observations at the model-based hierarchical stage of clustering, to speed up the algorithm. 
This is easily done by setting the argument \code{samp = TRUE}, and by specifying the number of observations to be used in the hierarchical clustering subset with \code{sampsize}.

Consider the following simulation scheme which constructs a medium sized dataset on five dimensions. Only the first two variables contain clustering information, the third is highly correlated with the first one, whereas the remaining features are simply noise variables.

\begin{CodeInput}
R> library(MASS)
R> n = 1000     # sample size
R> pro = 0.5    # mixing proportion
R> mu1 = c(0,0) # mean vector for the first cluster
R> mu2 = c(3,3) # mean vector for the second cluster
R> sigma1 = matrix(c(1,0.5,0.5,1),2,2)       # covar matrix for the first cluster
R> sigma2 = matrix(c(1.5,-0.7,-0.7,1.5),2,2) # covar matrix for the second cluster
R> X = matrix(0, n, 5); colnames(X) = paste("X", 1:ncol(X), sep ="")
R> # generate the clustering variables
R> set.seed(123)
R> u = runif(n)
R> Class = ifelse(u < pro, 1, 2)
R> X[u < pro, 1:2]  = mvrnorm(sum(u < pro), mu = mu1, Sigma = sigma1)
R> X[u >= pro, 1:2] = mvrnorm(sum(u >= pro), mu = mu2, Sigma = sigma2)
R> # generate the non-clustering variables
R> X[,3] = X[,1] + rnorm(n)
R> X[,4] = rnorm(n, mean = 1.5, sd = 2)
R> X[,5] = rnorm(n, mean = 2, sd = 1)
R> clPairs(X, Class, gap = 0.2)          
\end{CodeInput}
% dev.copy2pdf(file = "sim1_fig1.pdf", width = 9, height = 8)

\begin{figure}
\centering
\includegraphics[width=0.8\linewidth]{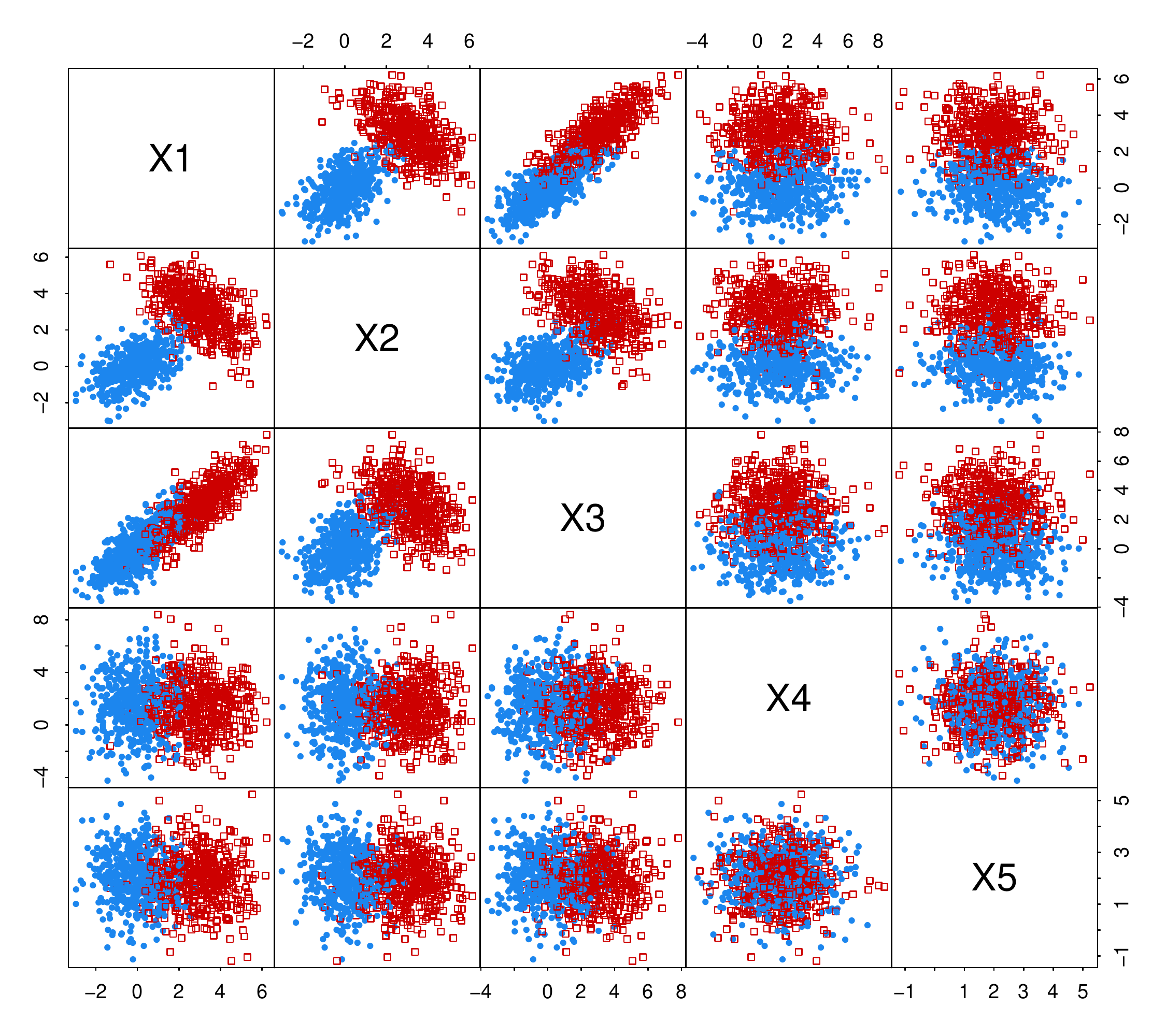}
\caption{Scatterplot matrix of simulated data with points marked according to the known groups.}
\label{fig1:sim1}
\end{figure}

We may compare the procedure which uses sampling at the hierarchical stage with the default call to \code{clustvarsel}, both in term of computing time, using the function \code{system.time}, and in term of clustering accuracy.
\begin{CodeInput}
R> system.time(result1 <- clustvarsel(X, G = 1:5, samp = TRUE, sampsize = 200))
\end{CodeInput}
\begin{CodeOutput}
   user  system elapsed 
  6.033   0.009   6.044
\end{CodeOutput}
\begin{CodeInput}
R> system.time(result2 <- clustvarsel(X, G = 1:5))
\end{CodeInput}
\begin{CodeOutput}
   user  system elapsed 
  9.739   0.044   9.820 
\end{CodeOutput}

Thus, by using sub-sampling for the initial hierarchical clustering we get an algorithm that is  38\% faster with the same accuracy: 
\begin{CodeInput}
R> result1$subset
\end{CodeInput}
\begin{CodeOutput}
X2 X1 
 2  1 
\end{CodeOutput}
\begin{CodeInput}
R> result2$subset 
\end{CodeInput}
\begin{CodeOutput}
X2 X1 
 2  1 
\end{CodeOutput}

\begin{figure}
\centering
\subfloat[][system time vs sample size]{\includegraphics[width=0.48\linewidth]{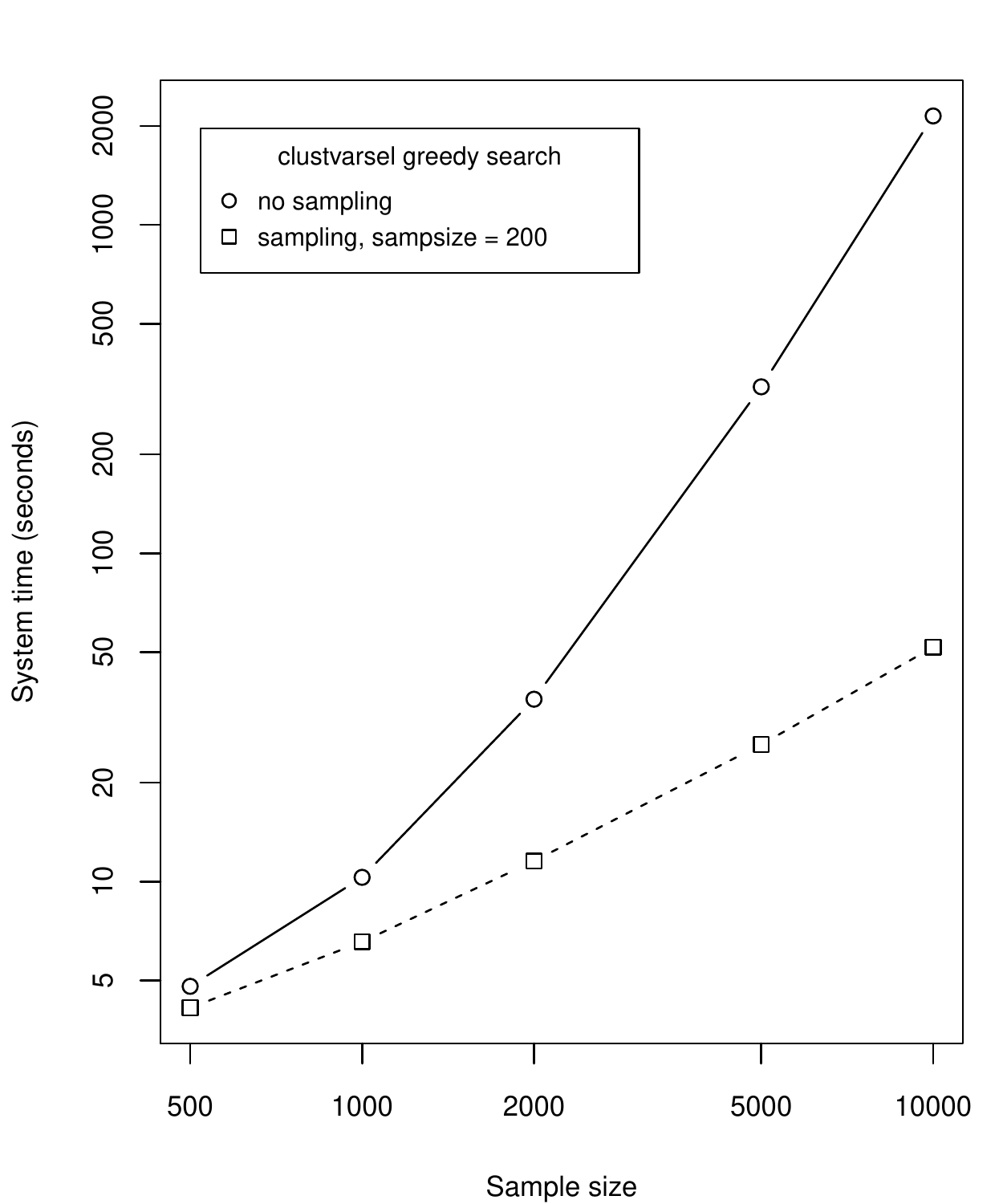}}
\quad
\subfloat[][relative gain of sub-sampling vs sample size]{\includegraphics[width=0.48\linewidth]{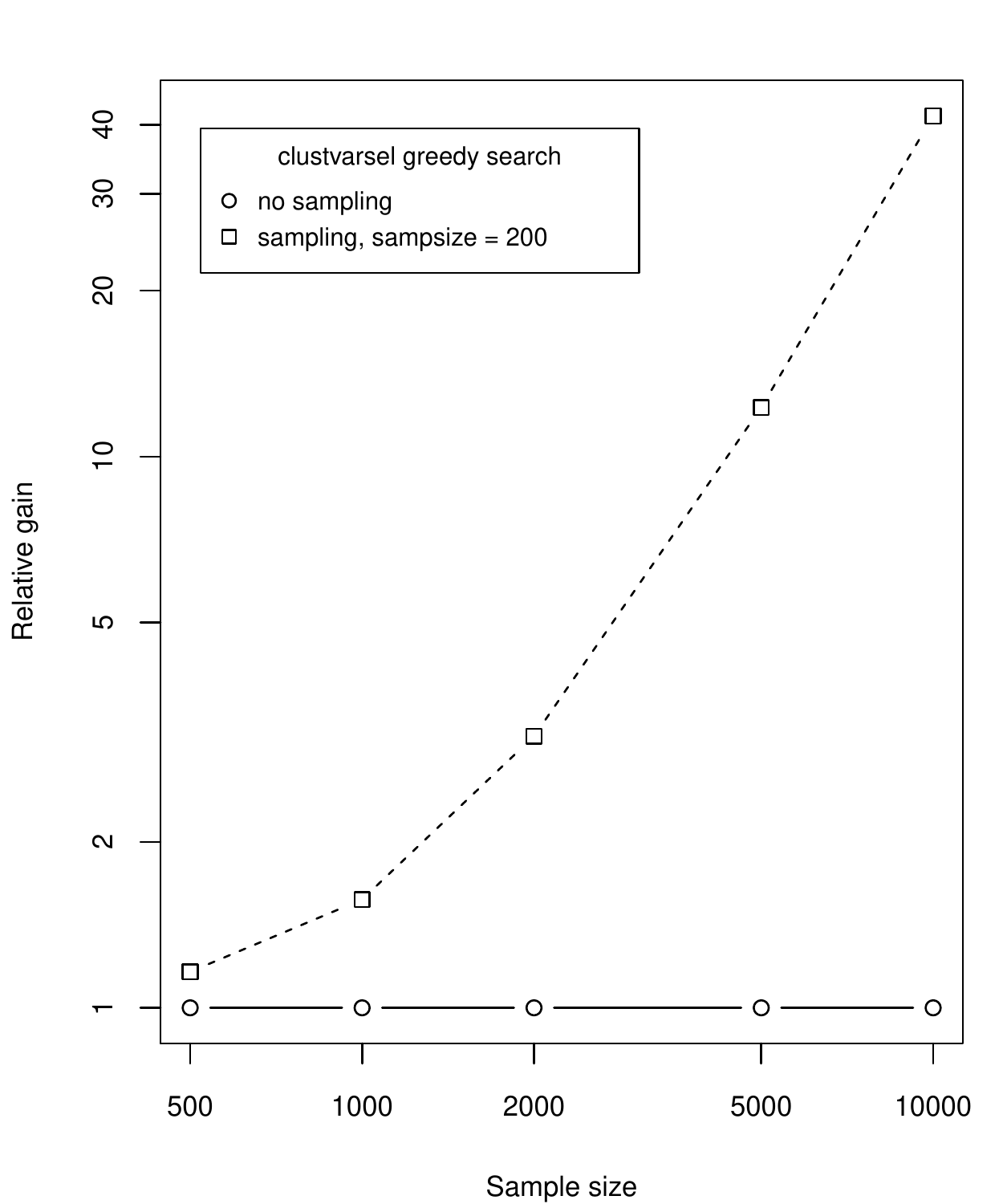}}
\caption[Computing time vs sample size for subsampling method]
{Comparison of computing time vs sample size. Panel (a) shows the average over 10 replications for \code{clustvarsel} using sub-sampling with fixed size at 200 observations, and no sampling. Panel (b) shows the relative gain of sub-sampling, calculated as the ratio of system times for no sampling and sub-sampling. All axes are on the logarithmic scale.}
\label{fig2:sim1}
\end{figure}

To investigate the effect of sampling as the number of observations increase we conducted a small simulation study by replicating the above simulation setting with different sample sizes and fixed size at 200 observations for choosing the initial starting points. Figure~\ref{fig2:sim1} shows the results averaged over 10 replications. 
Panel (a) reports the computing time required as the sample size grows, whereas panel (b) shows the relative gain from using a subset of observations at the initial hierarchical stage. As can be seen, efficiency improves roughly exponentially as the number of observations increases, with sampling being about 40 times faster at $10,000$ cases. As the system time required increases linearly for sampling, when no sampling is used at the initial stage the time required increases approximately exponentially. 
Furthermore, in all the replications the first two variables have been selected by both methods. Hence, the improvement in terms of computational efficiency has not caused any deterioration in terms of accuracy.

\subsection{Headlong search}
\label{sec:headlong}

When a dataset contains a large number of variables we may find that using the headlong search algorithm option (\code{search = "headlong"}) is faster than the default greedy search. To show an example we simulated a dataset analogous to the previous one for the clustering variables, then six more irrelevant variables were added, some correlated with the clustering ones, some independent and some correlated among themselves. Then, we may compare the time required by using the headlong method and using the greedy method.

\begin{CodeInput}
R> library(MASS)
R> n = 400      # sample size
R> pro = 0.5    # mixing proportion
R> mu1 = c(0,0) # mean vector for the first cluster
R> mu2 = c(3,3) # mean vector for the second cluster
R> sigma1 = matrix(c(1,0.5,0.5,1),2,2)       # covar matrix for the first cluster
R> sigma2 = matrix(c(1.5,-0.7,-0.7,1.5),2,2) # covar matrix for the second cluster
R> X = matrix(0, n, 10); colnames(X) = paste("X", 1:ncol(X), sep ="")
R> # generate the clustering variables
R> set.seed(1234)
R> u = runif(n)
R> Class = ifelse(u < pro, 1, 2)
R> X[u < pro, 1:2]  = mvrnorm(sum(u < pro), mu = mu1, Sigma = sigma1)
R> X[u >= pro, 1:2] = mvrnorm(sum(u >= pro), mu = mu2, Sigma = sigma2)
R> # generate the non-clustering variables
R> X[,3] = X[,1] + rnorm(n)
R> X[,4] = X[,2] + rnorm(n)
R> X[,5] = rnorm(n, mean = 1.5, sd = 2)
R> X[,6] = rnorm(n, mean = 2, sd = 1)
R> X[,7:8] = mvrnorm(n, mu = mu1, Sigma = sigma1)
R> X[,9:10] = mvrnorm(n, mu = mu2, Sigma = sigma2)
\end{CodeInput}

\begin{CodeInput}
R> system.time(result1 <- clustvarsel(X, G = 1:5, search = "headlong"))
\end{CodeInput}
\begin{CodeOutput}
   user  system elapsed 
  5.818   0.029   5.847 
\end{CodeOutput}

\begin{CodeInput}
R> system.time(result2 <- clustvarsel(X, G = 1:5))
\end{CodeInput}
\begin{CodeOutput}
   user  system elapsed 
 10.107   0.033  10.139 
\end{CodeOutput}

\begin{figure}
\centering
\subfloat[][system time vs sample size]{\includegraphics[width=0.48\linewidth]{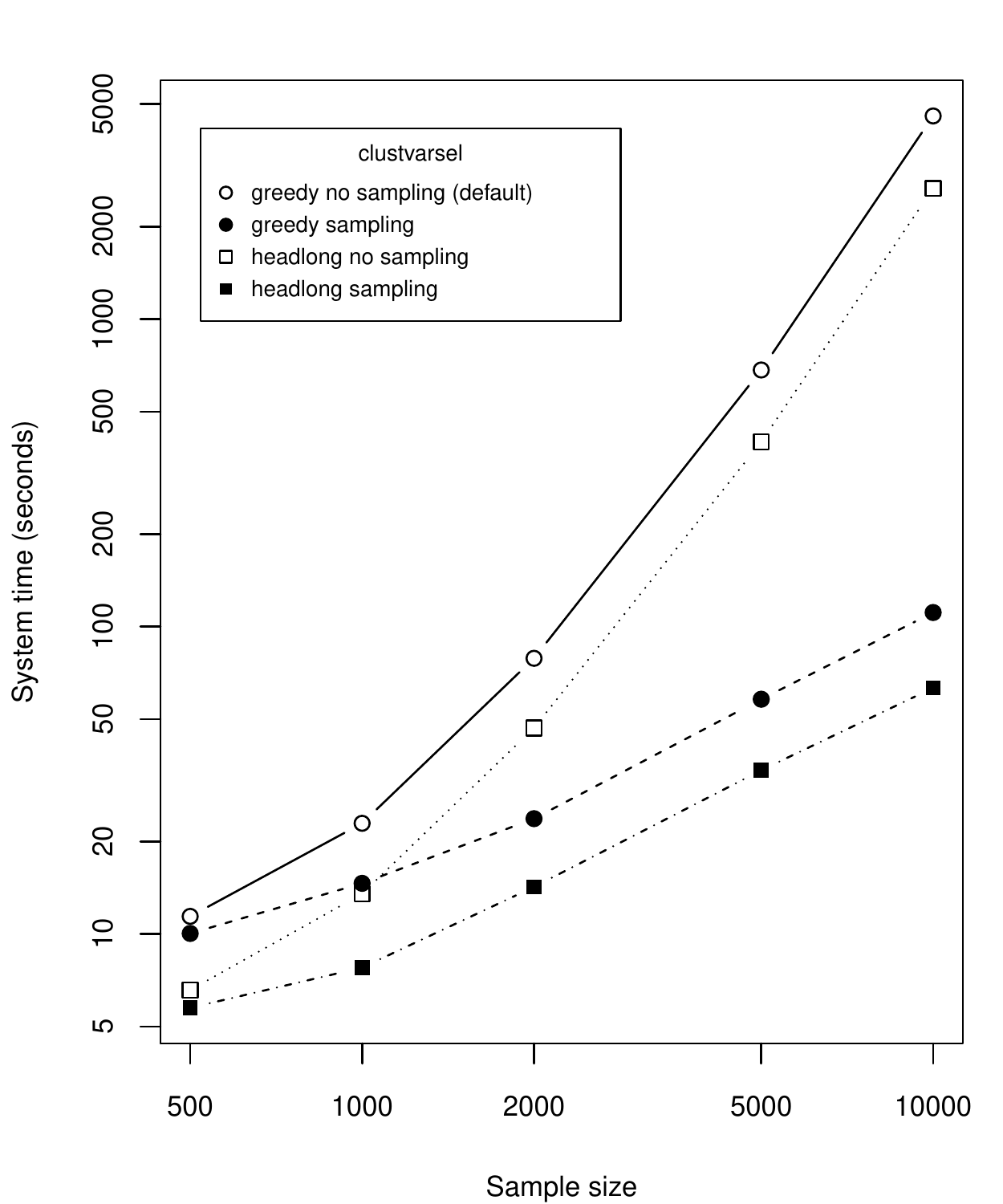}}
\quad
\subfloat[][relative gain of sub-sampling vs sample size]{\includegraphics[width=0.48\linewidth]{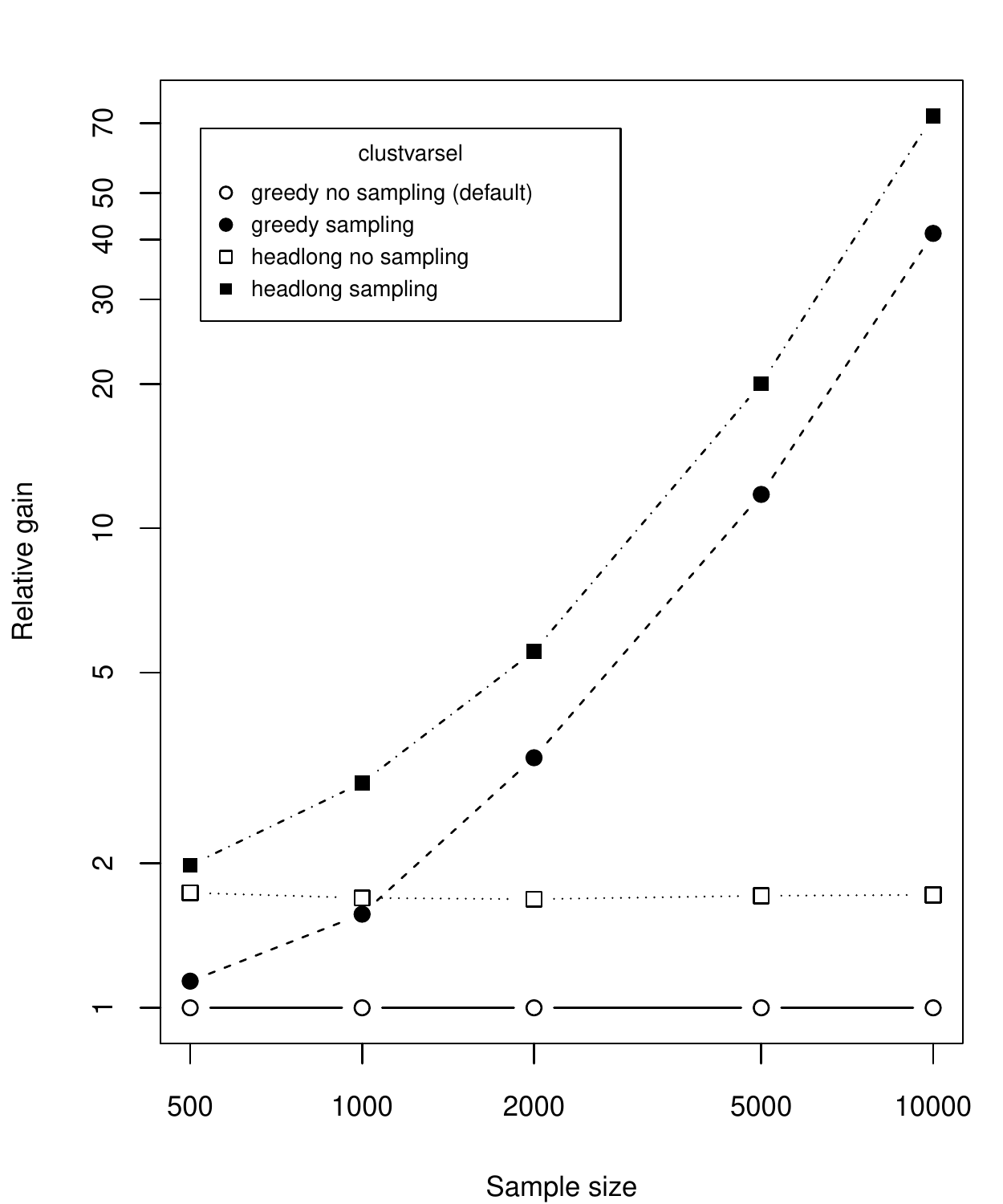}}
\caption[Computing time vs. sample size for greedy and headlong methods]
{Comparison of computing time vs sample size. Panel (a) shows the average over 10 replications for \code{clustvarsel} using \code{search = "greedy"} with and without sampling, and \code{search = "headlong"} with and without sampling. A fixed value \code{sampsize = 200} is used throughout. Panel (b) shows the relative gain, calculated as the ratio of system times, of each strategy against the default greedy search with no sampling. All axes are on the logarithmic scale.}
\label{fig1:sim2}
\end{figure}

In situations where there are many observations and a large number of variables, sub-sampling at the hierarchical initialisation step and the headlong search can be used concurrently to improve computational efficiency.
A small simulation study was conducted by replicating the previous simulation scheme with different sample sizes. The methods compared are greedy and headlong searches, without and with sampling using \code{sampize = 200}.
The results averaged over 10 replications are shown in Figure~\ref{fig1:sim2}. Without sampling, headlong search is faster than greedy search with a constant relative gain of about $1.7$. The use of sampling at the initial hierarchical stage enables us to achieve an exponential relative gain as the sample size increases for both type of searches. Also in this case, the headlong search appears to be about twice as fast as the greedy search. 

The speed/optimally tradeoff in a headlong search can be changed by increasing or decreasing the different levels, e.g. by setting the upper level to 10 instead of 0 we would require a variable to have stronger evidence of clustering before it is included, and by setting the lower level to 0 we would remove variables that at any stage have evidence of clustering weaker by any amount than evidence against clustering.

\subsection{Parallel computing}
\label{sec:parallel}

Parallel computing is a form of computation in which the required calculations are performed simultaneously, either on a single multi-core processors machine or on a cluster of multiple computers. 

Direct support of parallelism in \proglang{R} is available since version 2.14.0 (released in October 2011) through the package \pkg{parallel} \citep{Rpkg:parallel}. This is essentially a merger of the \pkg{multicore} package \citep{Rpkg:multicore} 
and the \code{snow} package \citep{Rpkg:snow}. 
The \code{multicore} functionality supports parallelism via forking, which is a concept from POSIX operating systems, so it is available on all \proglang{R} platforms except Windows.
In contrast, \pkg{snow} supports different transport mechanisms (e.g. socket connections) to communicate between the master and the workers, and it is available on all operating systems.
Other approaches to parallel computing in \proglang{R} are available as described in \citet{McCallum:Weston:2011}. For an extensive list of packages see CRAN task view on \textit{High-Performance and Parallel Computing with R} at \url{http://cran.r-project.org/web/views/HighPerformanceComputing.html}.

The greedy search discussed in Section~\ref{sec:methods} constitutes an embarrassingly parallel problem, i.e. one for which little or no effort is required to separate the problem into a number of parallel tasks. Essentially, the sequential evaluation of candidate variables for inclusion or exclusion, which is the most time consuming task, can be done in parallel.
For the actual implementation in \pkg{clustvarsel} we used the \pkg{doParallel} package \citep{Rpkg:doParallel},
a ``parallel backend'' which acts as an interface between the \pkg{foreach} package \citep{Rpkg:foreach} and the \pkg{parallel} package. Essentially, it provides a mechanism needed to execute for loops in parallel. 

To specify if parallel computing should be used in the evaluation of the $\BICdiff$ criterion in \eqref{eq:BICdiff}, the optional argument \code{parallel} must be set to \code{TRUE} in the \code{clustvarsel} function call. In this case all the available cores, as returned by the \code{detectCores} function, are used. A numeric value specifying the number of cores to employ can also be specified in the optional argument \code{parallel}. Finally, the parallelisation functionality depends on system OS: on Windows only 'snow' type functionality is available, whereas on Unix/Linux/Mac OSX both 'snow' and 'multicore' (default) functionalities are available.

As an example, consider a sample of $n=200$ observations generated according to the simulation scheme described in Section~\ref{sec:headlong}. We may compare the sequential greedy backward/forward search with a parallel version of the algorithm with the default maximum cores available and by specifying 2 cores (on a MacBook Pro with i5 Intel\textsuperscript{\textregistered} CPU with 4 cores running at $2.3$GHz and with $4$GB of RAM):

\begin{CodeInput}
R> system.time(result1 <- clustvarsel(X, G = 1:9, direction = "backward"))
\end{CodeInput}
\begin{CodeOutput}
   user  system elapsed 
103.860   0.509 104.774
\end{CodeOutput}

\begin{CodeInput}
R> system.time(result2 <- clustvarsel(X, G = 1:9, direction = "backward", 
                                      parallel = TRUE))
\end{CodeInput}
\begin{CodeOutput}
   user  system elapsed 
158.578   5.685  51.254 
\end{CodeOutput}

\begin{CodeInput}
R> system.time(result2 <- clustvarsel(X, G = 1:9, direction = "backward", 
                                      parallel = 2))
\end{CodeInput}
\begin{CodeOutput}
   user  system elapsed 
119.161   2.887  67.221
\end{CodeOutput}

In this case, by using 4 cores we were able to halve the computing time, whereas a 35\% speed up is achieved using 2 cores.

By using a machine with $P$ processors instead of just one, we would like to obtain an increase in calculation speed of $P$ times. As shown above, this is not the case because in the implementation of a parallel algorithm there are some inherent non-parallelizable parts and communication costs between tasks \citep{Nakano:2012}.
Amdahl's Law \citep{Amdahl:1967} is often used in parallel computing to predict the theoretical maximum speedup when using multiple processors. If $f$ is the fraction of non-parallelizable task and $P$ is the number of processors in use, then the maximum speedup achievable on a parallel computing platform is given by
\begin{equation}
S_P = \frac{1}{f + (1-f)/P} .
\label{eq:AmdahlLaw}
\end{equation}
In the limit, the above ratio converges to $S_{\max} = 1/f$, which represents the maximum increase of speed achievable in theory, i.e. by a machine with an infinite number of processors.

\begin{figure}[htb]
\centering
\includegraphics[width=0.7\linewidth]{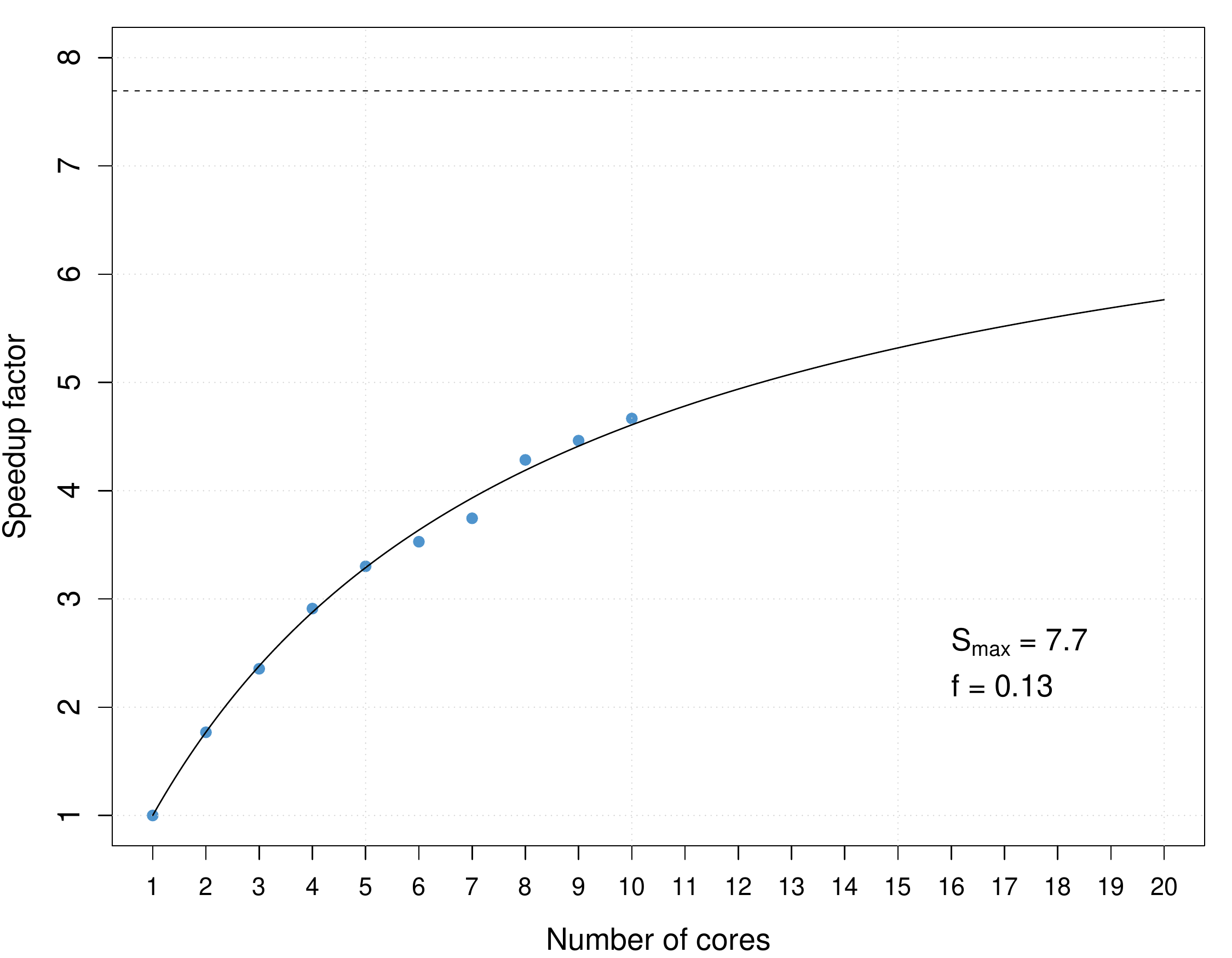}
\caption[Speedup by number of cores in the parallel algorithm]
{Graph of speedup factor vs the number of cores employed in the parallel algorithm for backward/forward subset selection in model-based clustering. The estimated fraction of non-parallelizable task is estimated as $f=0.13$, which gives a maximum speedup achievable by parallelisation of around $7.7$ times the sequential algorithm.}
\label{fig:AmdahlLaw}
\end{figure}

To investigate the performance of our parallel algorithm implementation, we conducted a small simulation study using the above simulation setting for increasing numbers of cores. The study was performed on a 24 cores Intel$^{\textregistered}$ Xeon\textsuperscript{\textregistered} CPU X5675 running at $3.07$GHz and with $128$GB of RAM.
Figure~\ref{fig:AmdahlLaw} shows the results averaged over 10 replications. The points represent the observed speedup factor (obtained as $s_P=t_1/t_P$, where $t_P$ is the system time employed using $P$ cores) for running the backward algorithm with up to 10 cores. The curve represents the Amdahl's Law \eqref{eq:AmdahlLaw} with $f$ estimated by non-linear least squares. It turns out that the estimated fraction of sequential part of the backward/forward search for variable selection is $\hat{f} = 0.13$, which yields a maximum speedup of $S_{\max} = 7.7$.

\section{Conclusions and future work}
\label{sec:discuss}

This paper has presented the \proglang{R} package \pkg{clustvarsel} which provides a convenient set of tools for variable selection in model-based clustering using a finite mixture of Gaussian densities. Stepwise greedy search and headlong algorithm are implemented in order to find the (locally) optimal subset of variables with cluster information. 
The computational burden of such algorithms can be decreased by some ad hoc modifications in the algorithms, or via the use of parallel computation as implemented in the package.
Examples illustrating the use of the package in practical applications have been presented.
Finally, given the vast solution space, other optimisation techniques could be usefully employed. For instance, the use of genetic algorithms as described in \citet{Scrucca:2014} will be included in a future release of the package. 

% References
\baselineskip=15pt
\bibliographystyle{chicago}
\bibliography{clustvarsel_arxiv}

\end{document}